\documentclass[prd,aps,onecolumn,floatfix,showpacs,preprintnumbers,amsmath,amssymb]{revtex4}
\usepackage{amsmath}
\usepackage{graphicx}
\usepackage{dcolumn}
\usepackage{bm}
\usepackage{amssymb}
\usepackage{latexsym}

\def\be{\begin{equation}}
\def\ee{\end{equation}}
\def\ba{\begin{eqnarray}}
\def\ea{\end{eqnarray}}
\def\ben{\begin{equation}}
\def\een{\end{equation}}
\def\bea{\begin{eqnarray}}
\def\eea{\end{eqnarray}}
\def\tg{\tilde{g}}
\def\tA{\tilde{A}}
\def\cT{{\cal T}}
\def\cD{{\cal{D}}}
\def\cO{{\cal{O}}}
\def\cA{{\cal{A}}}
\def\cB{{\cal{B}}}

\def\vx{{\vec{x}}}

\usepackage{color}

\begin{document}


\title{Fluctuations in a Cosmology with a Space-Like Singularity and their Gauge Theory Dual Description}

\author{Robert H. Brandenberger\footnote{Email: rhb@physics.mcgill.ca}}
\affiliation{Department of Physics, McGill University, Montr\'eal, QC, H3A 2T8, Canada}

\author{Yi-Fu Cai\footnote{Email: yifucai@ustc.edu.cn}}
\affiliation{CAS Key Laboratory for Researches in Galaxies and Cosmology,
 Department of Astronomy, University of Science and Technology of China, 
 Chinese Academy of Sciences, Hefei, Anhui 230026, China}
\affiliation{Department of Physics, McGill University, Montr\'eal, QC, H3A 2T8, Canada}

\author{Sumit R. Das\footnote{Email: das@pa.uky.edu}}
\affiliation{Department of Physics and Astronomy, University of Kentucky, Lexington, KY, 40506, USA}

\author{Elisa G. M. Ferreira\footnote{Email: elisa.ferreira@mail.mcgill.ca}}
\affiliation{Department of Physics, McGill University, Montr\'eal, QC, H3A 2T8, Canada}

\author{Ian A. Morrison\footnote{Email: imorrison@physics.mcgill.ca}}
\affiliation{Department of Physics, McGill University, Montr\'eal, QC, H3A 2T8, Canada}

\author{Yi Wang\footnote{Email: phyw@ust.hk}}
\affiliation{Department of Physics, The Hong Kong University of Science and Technology, Clear Water Bay, Kowloon, Hong Kong, P.R.China}

\pacs{98.80.Cq}

\begin{abstract}

We consider a time-dependent deformation of anti-de-Sitter (AdS) space-time which
contains a spacelike ``singularity" - a space-like region of high curvature.
Making use of the AdS/CFT correspondence we can map the bulk dynamics onto the boundary. The 
boundary theory has a time dependent coupling constant which becomes small
at times when the bulk space-time is highly curved. We investigate 
the propagation of small fluctuations of a test scalar field from early times before the bulk
singularity to late times after the singularity. Under the assumption that the AdS/CFT
correspondence extends to deformed AdS space-times, we can map the bulk
evolution of the scalar field onto the evolution of the boundary gauge field. The time 
evolution of linearized fluctuations is well defined in the boundary theory as long as the 
coupling remains finite, so that we can extend the boundary perturbations
to late times after the singularity. Assuming that the spacetime in the future of the 
singularity has a weakly coupled region near the boundary,  we reconstruct the bulk fluctuations 
after the
singularity crossing making use of generic properties of boundary-to-bulk propagators. 
Finally, we extract the spectrum of the fluctuations at late times given some initial spectrum. We
find that the spectral index is unchanged, but the amplitude increases due to
the squeezing of the fluctuations during the course of the evolution.
This investigation can teach us important lessons on how the spectrum of cosmological
perturbations passes through a bounce which is singular from the bulk point
of view but which is resolved using an ultraviolet complete theory of quantum
gravity. 

\end{abstract}

\maketitle

\section{Introduction}

The AdS/CFT correspondence \cite{Malda} is a most promising proposal
for a non-perturbative definition of string theory. Thus, this
correspondence should also have important consequences for early
universe cosmology. In fact, over the years there have been several
proposals which address the meaning of cosmological singularities in
the dual field theory \cite{HH,CH1,DMNT,CHT}.
The general idea is the following: consider an asymptotically AdS
space-time which is contracting towards a curvature
singularity. According to the AdS/CFT dictionary, this may correspond
to a dual conformal field theory which lives on the boundary which is
in a nontrivial unstable state \cite{HH,CHT} or which has a time
dependent coupling \cite{CH1, DMNT} which becomes small when the bulk
singularity is reached. While the bulk theory cannot be used to evolve
further in time, it may be possible to track the time evolution in the
dual field theory in a controlled fashion. It is then not unreasonable
to assume that the dual field theory admits a continuation in
time beyond the time $t_B = 0$ when the bulk singularity occurs.

There are several motivations for this investigation. One of the motivations comes from cosmology.
Although the inflationary
scenario \cite{Guth} is the current paradigm of early universe cosmology and has
been quite successful phenomenologically, it faces conceptual challenges. In
particular, a robust embedding into string theory has proven to be elusive (see
e.g. \cite{Rudelius}).
At the same time, it has been realized that there are alternative cosmological
scenarios which are at the moment also in agreement with cosmological
data. One of these is the ``matter bounce'' scenario (see \cite{RHBmbRev}
for a recent review), a bouncing scenario which begins with a matter-dominated
phase of contraction during which the scales which we observe today with
cosmological experiments exit the Hubble radius. It was shown in \cite{Wands, FB}
the if fluctuations begin in their Bunch-Davies vacuum at past infinity, that
the growth of the fluctuations on super-Hubble scales converts the vacuum
spectrum into a scale-invariant one for scales exiting the Hubble radius during
the matter phase of contraction. Adding a small cosmological constant (of
magnitude similar to the one observed today) leads to a small red tilt
in the spectrum \cite{Yifu}. The observed spectrum of curvature fluctuations
is indeed scale-invariant with a small red tilt (see \cite{Planck} for the most
recent data). 

In the context of effective field theory and Einstein gravity, it is difficult to obtain
a non-singular bouncing cosmology. One either needs to postulate that matter
violates the ``Null Energy Condition'' (NEC) during the bounce, or one needs to go
beyond Einstein gravity. Examples of the former are adding ghost condensate
matter \cite{ghost} or Galileon matter \cite{Galileon}, an example of the latter
is Horava-Lifshitz gravity in the presence of non-vanishing spatial curvature
\cite{HLbounce}. However, it is doubtful whether any of these constructions actually
can emerge from an ultraviolet complete theory such as string theory. Hence,
it would be very interesting to investigate if the AdS/CFT correspondence leads to a consistent bouncing cosmology \footnote{There are other approaches
to string theory which indicate the possibility of obtaining non-singular
bouncing cosmologies. One example is ``string gas cosmology'' \cite{BV} in
which the universe begins in an emergent high temperature stringy Hagedorn
phase, and in which the thermal string fluctuations in the Hagedorn phase
lead to a scale-invariant spectrum of fluctuations with a small red tilt \cite{NBV}.
Another example is the ``S-brane bounce'' of \cite{Costas}, in which an S-brane
arising at an enhanced symmetry point in the early universe leads to the
violation of the NEC which makes a non-singular bouncing cosmology
possible.}.

Regardless of the above motivation, it is clearly interesting to invstigate what 
happens to classical spacelike singularities in a complete theory of gravity. In 
particular, does the holographic correspondence predict a time evolution 
beyond this "singularity" ? 
Despite a lot of effort, it is not clear whether any of the AdS/CFT models
which contain {\em true singularities} in the bulk admit a smooth time
evolution in the dual theory. In the original model proposed in
\cite{CHT} which was based on earlier work of \cite{HH}, there were
some technical problems which indicated that the time evolution past
the singularity was not under control \cite{CHTproblem}, mainly due to
the back-reaction of the fluctuations on the background
space-time. There have been attempts to overcome these obstacles
\cite{CHT2, ST}, but the final verdict is
still out.

In the works of \cite{CH1,CH2,DMNT,DMNT2,DMNT3} a bulk dilaton field $\phi$ had a time
dependent (or a null coordinate dependent) boundary condition so that
$e^\phi$ becomes small at some time (or null time), while Einstein
frame curvatures become large in the bulk, signifying a singularity. 
When the singularity is null, the dual theory appears to predict a
smooth time evolution, and because of the absence of particle
production one expects that the spacetime is smooth in the future.
However for backgrounds with space-like singularities, as in \cite{DMNT3}
there is no clear conclusion. Even though the background supergravity
solution is time symmetric, the issue relates to the effect of
fluctuations. In the boundary field theory, the question becomes that
of particle production. In \cite{DMNT3} it was argued that in the case
when the boundary theory coupling hits a zero, the time evolution of
each individual momentum mode is in fact singular. However it was not
clear what happens when one considers the full field theory. 
In a regulated version of the theory where the boundary coupling becomes 
small but does not hit a zero, time evolution is well defined.
However, the energy due to particle production at times
after the crunch would be large, and the spacetime will not bounce
back to pure AdS even at very late times. On general gounds, one 
might expect that a black brane is formed \cite{DMNT3} \footnote{
There are AdS cosmologies in
global AdS where the coupling enters a weak coupling region {\em
  slowly} where \cite{DMNT4} argue that the time evolution is
smooth. In this model, the Einstein frame curavatures are always
small, but string frame curvatures become large. Even though the dual
theory predicts a smooth evolution, the space-time beyond the crunch
cannot be determined reliably, though the energetics imply that big
black holes are not formed.}.

In this paper, we turn to a different aspect of the kind of
backgrounds with spacelike singularities studied in
\cite{DMNT,DMNT2,DMNT3} as a result of a time dependent boundary
condition for the bulk dilaton. The goal of our study is to
include cosmological perturbations in this picture. This is important for at least two
reasons. Firstly, it is important to study whether the background is stable against
the addition of fluctuations. Secondly, most of the data which we would like to
explain in cosmology concerns fluctuations (inhomogeneities in the distribution of
galaxies and anisotropies in the temperature of the cosmic microwave
background).

We will not explore further the important
question whether there are apparently singular AdS cosmologies which
lead to a bounce to a relatively empty space.
Rather, we will study models of the same type as 
\cite{DMNT,DMNT2,DMNT3} which have Kasner singularitites. However we
will keep $e^\phi$ finite (but small)
at all times, rather than going to zero, by putting in a cutoff in
time, $\xi$. More specifically, instead of the exact Kasner behavior
near the singularity
\ben
e^\phi(t)  = |t|^\alpha
\label{1-1}
\een
we will use 
\ben
e^\phi (t) = |t|^\alpha \theta (|t| - \xi) +
\frac{\cosh^2(M\xi)}{\cosh^2(Mt)}\xi^{\alpha}\theta (\xi - |t|)
\label{1-2}
\een
We do
not know of exact solutions with such a cutoff dilaton : we
{\em assume} that these can be constructed. In the presence of such a
finite cutoff, the boundary gauge theory is well defined : our aim to
study some aspects of this. In some sense, the spirit of our
investigation is similar to that of \cite{HHnew} and \cite{HHNewrefs}
(see also \cite{otherholo}) where signatures of a past Kasner singularities in the dual field
theory were studied.

We want to determine how the {\em spectrum} of
fluctuations evolves as the system passes through the "singularity".
This question is independent of nature of the late time space-time so long as 
there is a region of normal spacetime near the horizon, which is where the 
fluctuations are measured. 

This question is particularly interesting if any of these models reliably predict a bounce
since in this case connections with bouncing cosmologies studied by
many cosmologists can be made. There are classes of scenarios
where, starting from vacuum perturbations at early times in the
contracting phase, a scale-invariant spectrum 
of fluctuations is generated before the bounce. This occurs
both in matter bounce scenarios \cite{Wands, FB} for scales
which exit the Hubble radius in the matter-dominated phase of
contraction, and also in Ekpyrotic models (in the presence of
entropy fluctuations) \cite{Ekp, NewEkp}. To obtain a connection
with observations, the spectrum after the bounce needs to be
determined. It has been
shown \cite{Durrer} that the form of the spectrum after the bounce can
depend on details of the bounce, although in many toy models one finds
that on large scales the spectrum is preserved (see e.g. the analysis
of \cite{bounceflucts}). A result concerning the transfer of fluctuations
in an ultraviolet complete theory is thus highly desired.  However, this question 
is also of interest for the particular model which is analyzed in this paper, which 
in all likelihood produces a black brane.

As is clear from previous work, the gauge theory is in a highly excited
state as one approaches the region of weak 't Hooft coupling, possibly in a coherent state. In such
a situation, we expect that we can learn a lot from the {\em classical
  limit} of the Yang-Mills theory. We therefore study the time
evolution of small fluctuations around the background across the
region of weak coupling.  Such small fluctuations are related,
by the AdS/CFT correspondence to bulk fluctuations. The kind of
fluctuations we are interested in are those which are given by
correlation functions on a fixed radial slice on AdS, close enough to
the boundary. As we discussed, the space time in the future might
contain black holes (branes). However so long as the space-time near
the boundary is smooth and normal, we can use the AdS/CFT dictionary
to translate boundary fluctuations to bulk fluctuations.

In this paper we take a preliminary step towards computing the transfer
of cosmological fluctuations from before the beginnig to after the
end of the high curvature bulk regime. 
We begin with a given spectrum of cosmological perturbations in the
contracting phase of the bulk, while the bulk is still weakly coupled.
At the time when the bulk becomes strongly coupled (and,
correspondingly, the boundary conformal field theory becomes weakly
coupled) we map the fluctuations onto fluctuations of the gauge fields
in the boundary theory. We then evolve the fluctuations to the future
in the weakly coupled region on the boundary. In our classical
approximation, it is now straightforward to find the fate of these
fluctuations at late times. The third step of our analysis is the
reconstruction of bulk fluctuations from the boundary data to the
future of the bulk singularity. Note that we are interested in
fluctuations on scales of current cosmological interest. These scales
are infrared modes from the point of view of the physics which we are
considering. Specifically, the wavelength of the modes we are
interested in is larger than the Hubble radius at the times between $-
t_b$ and $t_b$ when we evolve the fluctuations on the boundary.

Our main result is that the momentum dependence of classical fluctuations for
momenta much smaller than the cutoff ($k\xi \ll 1$) does not change after
crossing the weak 't Hooft coupling region, while their amplitudes
change by $O(1)$ factors. This means that the spectrum of
fluctuations of the dilaton field near the boundary also have this
behavior. In particular, if we start out with a nearly scale invariant
spectrum with a red tilt (as is the case for models of matter bounce),
the spectrum right after the ``bounce'' will remain the same.
While this result is shown for the bulk dilaton, 
we conjecture that the result
we obtain will directly apply to the evolution of the gravitational wave
spectrum, the reason being that a test scalar field in a cosmological
background obeys the same equation of motion as that of the amplitude
of a particular polarization state of gravitational waves. 

This paper is organized as follows. In Section II, we review the 
proposed generalization of the AdS/CFT correspondence to
a time-dependent background \cite{DMNT, CH1}. Section III
consider a bulk scalar field and its dual in the boundary gauge
theory. As it turns out, the time-dependence of the background
scalar field induces a time-dependence of the mass of the
gauge field. In Section IV we study the evolution of the gauge
field given the time-dependence of the coupling constants
induced by the non-trivial scalar field in the bulk. Since we
are interested in eventually computing linear fluctuations in the
bulk, we will focus on linear perturbations of the boundary field.
This leads to a dramatic simplification of the analysis. We
can work in Fourier space. Each Fourier mode obeys an
ordinary differential equation which is analogous to the equation
which cosmological fluctuations in a time-dependent background
obey in the context of standard General Relativistic perturbation
theory. Hence, we can use the accumulated knowledge about the
evolution of cosmological fluctuations in time-dependent backgrounds
to solve for the evolution of the linear boundary gauge field
perturbations through the time point $t = 0$ where the bulk
theory becomes singular. Since the boundary theory is weakly
coupled near the bulk singularity, the computations done in
the context of the boundary theory remain under controle. At
large positive times (when the bulk theory becomes weakly
coupled) we then reconstruct the bulk scalar field using boundary-to-bulk
propagators. This is discussed in Section V where we also
extract the power spectrum of the scalar field fluctuations
at late times and relate it to the initial spectrum before the
bulk singularity. We discuss and summarize our results in the
final section.

\section{Time-Dependent AdS Background and CFT Dual}

The original Maldacena conjecture is a duality between a Type IIB string theory 
on $AdS_5 \times S_5$ and a conformal field theory, a supersymmetric Yang-Mills 
(SYM) ${\cal N} = 4$ large $N$ $SU(N)$ gauge theory, living on the boundary of 
$AdS_5$ \cite{Malda}. 
The two dimensionless quantities on the bulk side are $R/l_s$, where $R$ stands for 
the AdS radius and $l_s$ is the string length and the string coupling constant $g_s$. 
These are related to the two dimensionless quantities in the SYM theory, the 
Yang-Mills coupling $g_{YM}$ and the rank of the gauge group $N$ by
\ben
\frac{R^4}{l_s^4} = 4\pi g_{YM}^2 N~~~~~~~g_s = g_{YM}^2
\label{2-1}
\een
The string coupling $g_s$ is of course given by the bulk dilaton field $\varphi$
\ben
g_s = {\rm exp} \langle \varphi \rangle
\label{2-2}
\een
There are two particularly noteworthy aspects of this correspondence.
Firstly, it relates a gravitational theory (the bulk theory) to a non-gravitational
field theory on the boundary. From this point of view, the challenge
of quantizing gravity suddenly takes on a completely new view. Secondly,
the duality is a strong coupling - weak coupling duality. The bulk
dilaton provides both a measure of the couplings in the bulk and
in the boundary. However, it is precisely when the bulk theory becomes
strongly coupled that the boundary theory becomes weakly coupled. In particular 
we can consider $N \gg 1$ so that bulk quantum effects are small, but 
$g_{YM}^2 N  \ll 1$ so that the boundary theory is weakly coupled - this would 
correspond to bulk curvature scales of the order of string scale, signifiying that 
stringy effects become important in the bulk.   This can be achieved, e.g. by 
having $g_s = g_{YM}^2 \ll 1/N$ for some fixed large $N$.

This suggests an interesting avenue to address the question of resolution of 
cosmological singularities. One possible way to do this is to consider a time 
dependent boundary condition of the bulk dilaton \cite{DMNT,CH1}. By the standard 
AdS/CFT correspondence the dual field theory living on the boundary now has a 
time dependent coupling. A cosmological singularity corresponds to a divergence 
of the gravitational coupling and thus to a region where conventional 
approaches to quantizing gravity will fail. However, by the AdS/CFT correspondence 
the bulk theory is dual to a non-gravitational theory on the boundary, and the bulk singularity 
corresponds to a point in time when the boundary theory becomes weakly coupled. 
Thus, the usual principles of field theory 
quantization should be applicable, and one has to deal with a weakly coupled 
field theory which, however, is time dependent.

The first step is to consider the low energy limit of the bulk
Type II string theory, namely Type II supergravity, and to focus on the
bosonic sector of this theory. The second step is to allow for a time
dependence in the bulk fields. The bulk fields involve the ten-dimensional
metric, the dilaton $\varphi$ and a five form $F_5$. The ansatz for such 
solutions is given by a non-trivial metric of the form
\ben
ds^2 = \frac{R^2}{z^2} [ dz^2 + \tg_{\mu\nu}(x^\mu)dx^\mu dx^\nu ] + R^2 d\Omega_5^2
\label{2-2}
\een
where $d\Omega_5^2$ is the standard metric on a unit 5-sphere. The surface $z=0$ is the 
boundary of the space-time. Note that the metric on a constant $z$ slice is a function of 
the coordinates $x^\mu$ only. This also holds for the
dilaton field $\varphi (x^\mu)$. Finally, the five form is given by
\ben
F_{(5)} = \omega_5 + \star_{10} \omega_5
\label{2-3}
\een
As shown in \cite{DMNT}, these bulk fields satisfy the full ten dimensional equations of motion
provided 
\be \label{EE}
{\tilde R}_{\mu \nu} \, = \, \frac{1}{2} \partial_{\mu} \varphi \partial_{\nu} \varphi \, ,
\ee
where ${\tilde R}_{\mu \nu}$ is the Ricci tensor of the metric ${\tilde g}_{\mu \nu}$,
and where the dilaton $\varphi$ obeys the Klein-Gordon equation
\be \label{KGE}
\partial_{A} \bigl( \sqrt{- {g}}  g^{A B } \partial_{B} \varphi \bigr) \, = \, 0 \, .
\ee
In the above, $g$ is the determinant of the full metric ${g}_{A B}$, and the indices
$A, B$ run over all five space-time dimensions. 

$AdS_5 \times S_5$ is of course a special case of a space-time described by the metric
(\ref{2-2}). In this case, the metric $\tg_{\mu\nu} = \eta_{\mu\nu}$, the Minkowski metric, and the dilaton is constant.
The radial coordinate $z$ runs from 
$z = 0$ at the boundary to $z = \pm \infty$ at the Poincar\'e horizons.

Time-dependent deformations of this background were considered in
\cite{DMNT, CH1, DMNT2, DMNT3, DMNT4, CH2}. Specifically, we shall consider a 
background of the form \cite{DMNT3} obtained by introducing a time-dependent
dilaton background and adjusting the metric of  ${\tilde g}_{\mu \nu}$ such
that the Einstein equation (\ref{EE}) and the Klein-Gordon equation (\ref{KGE})
remain satisfied. We will mainly deal with a solution where the
the boundary metric looks like that of a
Friedmann universe
\begin{eqnarray} \label{metric3}
 \tilde{g}_{\mu\nu}dx^{\mu}dx^{\nu} \, = \, -dT^2+a^2(T)\delta_{ij} dx^idx^j~,
\end{eqnarray}
with $T$ being the cosmic time and $a(T)$ the scale factor. The dilaton and
scale factor were taken to be
\begin{eqnarray}\label{sol_bg}
 a \sim |T|^{1/3} ~,~~ \varphi = \frac{2}{\sqrt{3}} {\rm ln}{\frac{|T|}{R}}~.
\end{eqnarray}
This corresponds to an early contracting phase leading to a ``Big Crunch''
singularity at $ T= 0$ followed by an expanding cosmology. In conformal time $t$, 
the four dimensional part of the metric is 
\begin{eqnarray} \label{metric33}
 \tilde{g}_{\mu\nu}dx^{\mu}dx^{\nu} \, = \, 2|t| \big[ -dt^2+\delta_{ij} dx^idx^j \big],
\end{eqnarray}
while the dilaton is given by
\ben
 \varphi = \sqrt{3}{\rm ln}{\frac{|t|}{R}}~.
\label{sol_bg2}
\een
If the singularity can be resolved by mapping the dynamics
to the boundary, we will have a stringy realization of a bouncing
scenario, though one would not expect a perfect bounce. As mentioned 
in the introduction we do not know yet if this indeed happens.

The dilaton profile (\ref{sol_bg2}) leads to a diverging string coupling at early and 
late times : this would seem to require incorporation of bulk quantum corrections. 
However it turns out that one can obtain solutions which have bounded values of 
the dilaton at all times, and whose behavior near the "singularity" is identical to the 
above. Such a solution is given by  the metric \cite{DMNT2}
\ben
ds^2 = \frac{R^2}{z^2} \big[ dz^2 + |\sinh(2t)| \{ -dt^2 + \frac{dr^2}{1+r^2}+r^2 d\Omega_2^2 \} \big]
\label{2-4}
\een
and a dilaton
\ben
e^{\varphi (t)} = g_s |\tanh (t/R)|^{\sqrt{3}} \, .
\label{2-5}
\een
Near $ t = 0$ the dilaton profile goes over to that in (\ref{sol_bg2}).  The metric (\ref{2-4}), 
whose four dimensional part is a FRW metric with constant negative curvature does not, 
however, become (\ref{metric33}) as $t \rightarrow 0$. The difference between the two, 
however, become increasingly unimportant as $t \rightarrow 0$ where the stress tensor is 
dominated by the time derivative of the dilaton rather than the spatial curvature.

A sketch of the space-time we are considering is given in Fig. 1. The vertical axis at $z = 0$ 
is time, the horizontal axis at $t = 0$ represents the $AdS$ radial coordinate. The
Poincar\'e horizons are at 45 degrees. While we will consider the solution (\ref{sol_bg2}) 
and (\ref{metric33}) it is useful to think of this as embbeded in the solution (\ref{2-5}) and 
(\ref{2-4}) with a bounded dilaton. Then at early times the Yang Mills coupling is 
$g_{YM}^2 = g_s$ and we will always consider $g_s \ll 1, N \gg 1$ with $(g_s N) \gg 1$. 
Thus the early time evolution is governed by the bulk supergravity equations.  The time 
$t_b$ is defined as the time when the 't Hooft coupling of the boundary theory becomes 
$O(1)$, i.e.
\ben
t_b/ R \sim (g_s N)^{-1/\sqrt{3}} \, .
\label{2-6}
\een
Thus, in the region $t > - t_b$ and 
$t < t_b$ the bulk gravity theory is weakly coupled.  For $ -t_b < t < t_b$ the bulk 
curvatures grow large and stringy effects becomes important, while the  boundary 
field theory is weakly coupled, although in the presence of a time dependent coupling. 

Note that the solutions considered above have a non-trivial boundary value of the dilaton 
and a boundary metric which is conformal to flat space. However there are no subleading 
normalizable pieces of these bulk fields. From the dual field theory point of view, it thus 
appears that we have non-trivial time dependent sources, but a trivial response. This 
indicates that such cosmologies correspond to some non-trivial initial states. However 
the nature of this state in terms of the gauge theory variables is not known.

There is another feature of these solutions which deserves mention. The singularity at 
$t = 0$ extends from the boundary to the bifurcation point of the Poincare horizon at 
$z=\infty$. This means that there is in fact a singularity at any finite Poincare time 
which is infinitely far from the boundary. Presumably this feature has something to 
do with the nature of the initial state. However, as pointed out in \cite{HHnew}, this 
bifurcation point singularity can be resolved by lifting the solution to one higher 
dimension and embedding the solution in a higher dimensional soliton solution.

In the following sections we are interested in evolving spectator massless bulk 
scalar field (which will be in fact taken to be the dilaton itself) 
perturbations from the past weakly coupled bulk region to the future
weakly coupled bulk region by mapping the state onto the boundary
at time $t = - t_b$, evolving on the boundary to positive times, and
using boundary-to-bulk reconstruction techniques to recover fluctuations
in the bulk. 

\begin{figure*}[t]
\begin{center}
\includegraphics[scale=0.6]{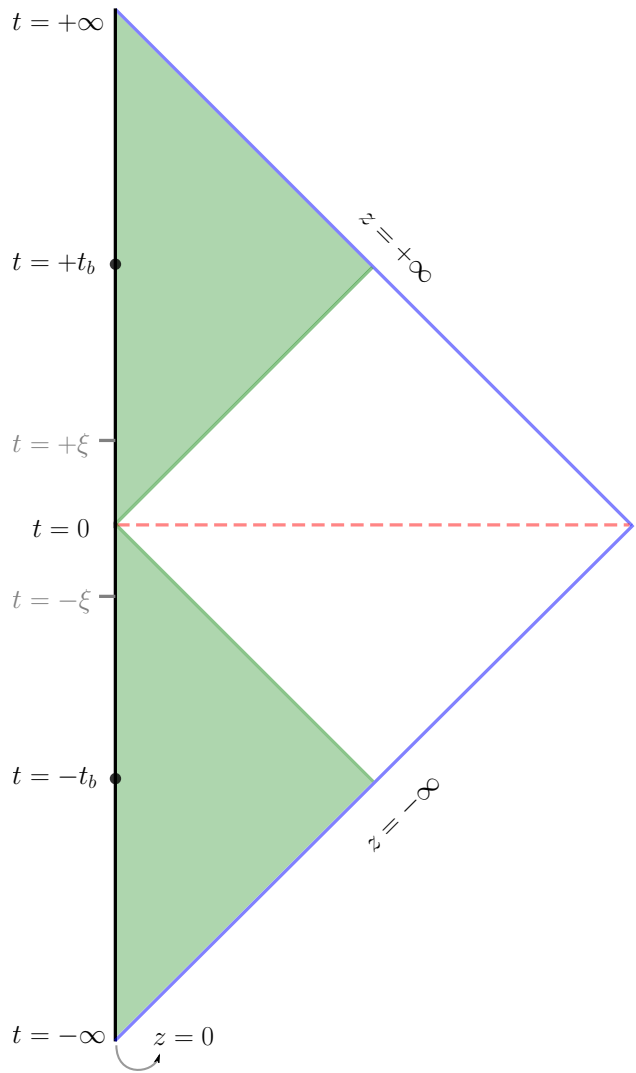}
\caption{Conformal diagram of the background space-time.
The vertical axis at $z = 0$ is time $t$, the horizontal 
direction at $t = 0$ represents the coordinate $z$. The
lines at 45 degrees are the Poincar\'e horizons at
$z = \pm \infty$. If there were no deformation of $AdS$, the 
region drawn would correspond to the Poincare patch of $AdS$.
We are considering a deformed space-time in which
the bulk gravity is strongly coupled between $t = - t_b$
and $t = t_b$, and singular at $t = 0$. At the same time the 
boundary gauge theory becomes weakly coupled for $t$ between
$- t_b$ and $t_b$. Hence, after the time $- t_b$ the evolution 
on the boundary becomes tractable in perturbation theory. 
On the future side of the bulk singularity, the boundary 
theory remains tractable perturbatively until
the time  $t_b$ when the the bulk theory becomes weakly coupled
again at the cost of the boundary theory becoming strongly coupled. 
At that time we can reconstruct the bulk information (at least in 
the vicinity of the boundary) from boundary data 
(see e.g. \cite{Hamilton1, Hamilton2, Kabat, ian, Enciso}). As we
will see, to study the evolution of the boundary fluctuations we
need to impose a cutoff at $t = \pm \xi$.}
\end{center}
\end{figure*}

\section{The Bulk Theory of a Test Scalar in a Contracting Universe}

To begin with, we consider a test scalar field of mass $m$ living in the bulk 
space-time (in the following section we will take this test scalar
field to be the dilaton itself). Its action takes the form 
\begin{eqnarray}
 S_\varphi \, = \, -\int d^5x \sqrt{-g} (g^{MN}\partial_M\varphi\partial_N\varphi + m^2\varphi^2)~.
\end{eqnarray}
where $g^{MN}$ is the metric of the five-dimensional deformed $AdS$ space-time
\ben
ds_5^2 = \frac{R^2}{z^2} [ dz^2 + \tg_{\mu\nu}(x^\mu)dx^\mu dx^\nu ] 
\label{3-1}
\een
with the four dimensional part $\tg_{\mu\nu}$ given by (\ref{metric3}). 
Varying the action with respect to the scalar field yields the following equation 
of motion
\begin{eqnarray}
 \ddot\varphi + 3H\dot\varphi - {a^{-2}}\varphi_{,ii}  + m^2\frac{R^2}{z^2}\varphi 
 -\varphi_{,zz} + \frac{3}{z}\varphi_{,z} \, = \, 0~,
\end{eqnarray}
where $H=\dot{a}/a$ is the Hubble parameter, and a dot denotes the derivative with 
respect to cosmic time $T$.

Since we are interested in the spectrum of fluctuations in the three spatially flat coordinates 
$x^{i}$, we first of all extract the $x^{i}$ dependence by expanding in Fourier modes. The 
resulting differential equation is a partial differential equation in $T$ and $z$, and we make 
a separation of variables ansatz to separate the $T$ and $z$ dependence. More specifically, 
we write
\be
\varphi(T, z, x^i) \, = \, \cT(T) Z(z) X(x^i) \, ,
\ee
where the $X(x^i)$ are the spatial Fourier modes, i.e. solutions of
\be
\nabla^2 X + k^2 X \, = \, 0 \, ,
\ee
with solutions which are positive or negative frequency oscillations
in the three vector ${\bf x}$, where the rescaled temporal field
\be
{\tilde \cT}(t) \, \equiv \, a(t) \cT(t)
\ee
(with $t$ being conformal time defined via $dT = a dt$ obeys the equation
\be \label{Teq}
{\tilde \cT}^{\prime \prime} + \bigl( \omega^2 a^2 + k^2 - \frac{a^{\prime \prime}}{a} \bigr) {\tilde \cT} \, = \, 0 \, ,
\ee
(a prime denoting the derivative with respect to $t$)
and the radial function $Z(z)$ obeys the equation
\be
Z_{, zz} - \frac{3}{z} Z_{,z} + \bigl( \omega^2 - \frac{m^2 R^2}{z^2} \bigr) Z \, = \, 0 \, .
\ee
The separation constant $\omega$ plays the role of a temporal frequency.
The solutions of the radial equation are Bessel functions
\be
Z(z) \, = C_J z^2 J_{\nu}(\omega z) + C_Y z^2 Y_{\nu} (\omega z) \, ,
\ee
with
\be
\nu \, = \, \sqrt{4 + m^2 R^2}
\ee

The temporal equation (\ref{Teq}) takes on the familiar form of the
coefficient function of a comoving
Fourier mode of a massive scalar field 
(mass given by $\omega^2$) in an expanding background space-time
which undegoes cosmological squeezing  (the final term on
the left hand side of the equation). In the case $\omega^2 = 0$ it is also 
the equation of motion which gravitational waves in an expanding space obey
\cite{MFB, RHBrev}. In particular, in the case of infrared modes for which
$k^2$ is negligible, then close to the singularity the mass term is negligible
and the squeezing term dominates. In this paper we will, however, not be
evolving the fluctuations in the bulk until the singularity, but only to the
point in time when the bulk theory ceases to be weakly coupled. 
Then, we will map them onto the boundary and evolve them with the 
boundary equations near the bulk singularity. 

Our main result does not depend on the initial fluctuation spectrum. As a concrete 
example, however, we could e.g. take the initial spectrum to be scale invariant for 
modes whose wavelength is larger than the Hubble radius.
During the phase of contraction the Hubble radius 
is decreasing in comoving coordinates (see Fig. 2). This is motivated by the fact that in 
several models of "bounce" cosmology (e.g. matter bounce of \cite{Wands,FB} or 
Ekpyrotic \cite{Ekp} scenarios) an
initial vacuum spectrum (see e.g. \cite{BD})
on sub-Hubble scales gets converted to
a scale-invariant one once the scales exit the Hubble radius
and undergo squeezing.

We will thus take the initial power spectrum 
\be
P(k) \, \equiv \, k^3 |(\delta \varphi)(k)|^2
\ee
on super-Hubble scales to have index 
$n = 0$ (where we are using the convention for the
index used in the cosmology community for gravitational waves, namely
$P(k) \sim k^n$).

\begin{figure*}[t]
\begin{center}
\includegraphics[scale=0.4]{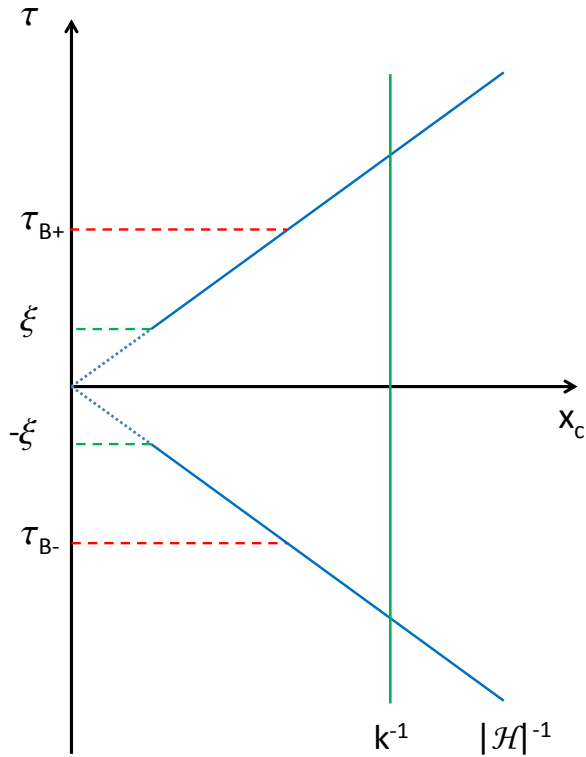}
\caption{Space-time sketch of the relevant times and length scales in
our deformed $AdS_5$. The horizontal axis is comoving spatial coordinate
(in direction perpendicular to the radial direction. The vertical direction is
conformal time. $\tau_{B+}$ and $\tau_{B-}$ are the times when the
coupling constant is $1$, the times $\pm \xi$ occur when the bulk curvature
reaches string scale. The vertical line represents the wavelength of a mode
which we are interested in. Note that the length is larger than the Hubble
radius at $\tau_{B+}$ and $\tau_{B-}$.  The blue lines at $45^o$ indicate
the Hubble radius.}
\end{center}
\end{figure*}

After discussing the bulk-to-boundary correspondence for
the fluctuations we will determine the initial spectrum of
perturbations of the boundary gauge fields which are induced
by the inhomogeneities of the bulk field discussed above.

\section{The Dual Boundary Theory}

\subsection{The Deformed Dual Boundary Theory}

For a pure $AdS$ bulk, the dual boundary theory is a ${\cal N}=4$ SYM theory.
According to the $AdS/CFT$ correspondence, fields in the bulk are related
to operators in the boundary theory. For example, the bulk metric $g_{\mu \nu}$
is dual to the boundary energy-momentum tensor $T_{\mu \nu}$. In this
paper we are interested in scalar field fluctuations in the bulk and their
boundary evolution. Specifically, we will take this scalar field $\varphi$ to be
the dilaton $\varphi$. In this case, the bulk scalar field is dual to
the trace of the square of the field strength tensor. For an exact $AdS$ bulk,
the boundary action is simply
\be \label{S_YM0}
 S_{\rm YM} \, = \, -\frac{1}{4}\int d^4 y \frac{1}{g_{\rm YM}^2} {\rm Tr}[F_{\mu\nu}F^{\mu\nu}] \, ,
 \ee
with
\be
 g_{\rm YM}^2 \, = \, e^\varphi \, = \, g_s \, .
\ee
Here, we denote the boundary coordinates by $y$.
We use a generalized $AdS/CFT$ correspondence according to which the
time-dependent dilaton in the bulk leads to a time-dependent gauge
coupling of the boundary theory, i.e. (\ref{S_YM0}) is generalized to be
\be \label{S_YM}
 S_{\rm YM} \, = \, -\frac{1}{4}\int d^4 y e^{- \varphi (t)} {\rm Tr}[F_{\mu\nu}F^{\mu\nu}] \, .
 \ee
The bulk solutions we are interested in have, in addition,  a boundary metric which 
is non-trivial and time dependent. However since this metric is conformally flat, and the 
boundary theory is the four dimensional SYM theory, the conformal factor decouples. 

What was said so far applies to an unperturbed theory. In the presence
of perturbations the correspondence becomes more involved \cite{Vijay}.
Each supergravity perturbation has two modes - one normalizable and
the other not. The normalizable mode determines the expectation value
of the boundary operator, the non-normalizable mode enters as the coupling of the 
operator in the boundary theory.

We are interested in evolving linearized bulk fluctuations. 
This means we will turn on normalizable bulk modes. 
Hence, we will
be evolving the linear fluctuations in the gauge field $A_{\mu}$ on
the boundary. The initial condiitions for the gauge field
fluctuations on the boundary are set by the dilaton perturbations
via the linearized version of the correspondence
\be \label{Operator}
\varphi (y)  \, \rightarrow \, \frac{1}{g_{\rm YM}^2}  \langle {\rm Tr}[F^2(y)] \rangle  \, ,
\ee
where the normalizable dilaton fluctuation behaves in the standard fashion as 
\ben
\varphi (z,y) \rightarrow z^4 \varphi (y)
\label{A-1}
\een
as one approaches the boundary $z = 0$.

The effects of linear dilaton fluctuations on the
evolution of the gauge field fluctuations would be a second order
effect in the amplitude of fluctuations. Hence, at linear level in
perturbation theory there is no such coupling and the only effect
of the dilaton on the gauge field fluctuations is via the dilaton-dependence of the gauge 
coupling constant at background level.

Given a spectrum of dilaton fluctuations at early times in the bulk,
we will evolve them in the bulk until bulk perturbation theory
breaks down at time $- t_b$. At that point, we compute
the boundary values of the dilaton fluctuations and use them
to determine the initial values of the gauge field fluctuations
$\delta A_{mu}$ at that time.

\subsection{Evolution of the Boundary Fluctuations Before the Singularity}

As we have mentioned, the dilaton field evolves as a function of cosmic time 
and initially takes a large value, and thus the Yang Mills coupling is also very 
large initially which implies that the boundary theory is strongly coupled in the 
far past. (It is useful to think in terms of the solutions (\ref{2-4}) and (\ref{2-5}) 
with bounded dilaton profiles). At these early times, the dynamics of the whole 
space-time can be studied making use of the bulk theory since the gravity sector 
is weakly coupled. As the universe contracts, the value of $e^\varphi$ decreases 
and the corresponding 't Hooft coupling of the field theory becomes $O(1)$ at the 
time $- t_b$ given by (\ref{2-6}).  This is the time where the gravity approximation 
begins to fail.

As the bulk space further contracts, the boundary theory becomes weakly 
coupled as $g_{\rm YM}^2 N <1$. 
When $t=0$ and $\varphi\rightarrow -\infty$, 
the bulk singularity point is reached. While it may appear that the boundary 
gauge theory is now free, as shown in \cite{DMNT3}, each momentum mode of the 
gauge theory displays a singular behavior. We therefore work with the modified 
dilaton profile (\ref{1-2}) with $\alpha = \sqrt{3}$ so that this is a genuine bulk 
solution for $|t| > \xi$. 

We will make a gauge choice
\ben
A_0 = 0 \, ,
\een
and also impose an additional constraint 
\ben
\partial^i A_i = 0 \, .
\een
The Gauss Law constraint is then automatically solved \cite{DMNT3}.

The form of the action (\ref{S_YM}) suggests a field redefinition (for an analysis in
terms of the original variables see Appendix A)
\ben
A_\mu \rightarrow \tA_\mu \equiv e^{-\varphi/2} A_\mu \, .
\een
This has two effects. The first is that it introduces a "mass" term for the redefined field. 
With a dilaton which depends only on time, the effective mass square is given by
\begin{eqnarray}\label{M_YM^2}
 M_{\rm YM}^2 \, = \, \frac{{\ddot \varphi}}{2} - \frac{{\dot{\varphi}}^2}{4} \, .
\end{eqnarray}
The second effect is to bring in a factor of $e^{\varphi / 2}$ in front of the cubic interaction 
term and a factor of $e^\varphi$ in front of the quartic interaction . This might suggest that 
as $t \rightarrow 0$ the nonlinear terms become small and can be ignored. However,  as 
shown in \cite{DMNT3} this is incorrect. If one substitutes a general solution of the linearized 
equations of motion (see below) into the action one finds that the fields $\tA_\mu$ blow up 
as $t \rightarrow 0$ in such a way that the original field $A_\mu$ becomes $O(1)$. Therefore, 
for arbitrary amplitudes the nonlinear terms cannot be ignored.
However we are interested in the time evolution of {\em small} fluctuations of the gauge field. 
The nonlinear terms are then suppressed because the amplitudes are small. In the following 
we will deal with the linear theory. Effects of nonlinearities will be explored in a future work.

Then the leading terms in the equation of motion for $\tilde{A}_i$
give
\begin{eqnarray}
 -\partial_\mu\partial^\mu \tilde{A}_i + M_{\rm YM}^2 \tilde{A}_i \, = \, 0~.
\label{B-0}
\end{eqnarray}
In the background \eqref{1-2} $M_{YM}^2$ becomes 
\ben
 M_{YM}^2(t) =
  \begin{cases}
    -\frac{\alpha(\alpha+2)}{4t^2}    & \quad \text{if } |t| > \xi\\
    - \frac{\alpha(\alpha+2)}{4\xi^2} & \quad \text{if } |t| < \xi \, .\\
  \end{cases}
\label{B-1}
\een
Note that the coefficients in the mass term diverge as $t \rightarrow 0$ if $\xi = 0$. This
is why the evolution of the fluctuations is non-trivial in spite of the fact
that the boundary gauge theory becomes weakly coupled at this time. Note that in
terms of the original variables $A_{\mu}$ there is no divergence. But a
branch cut in the solutions remains. Working with the rescaled variables
has the advantage that the equation of motion is similar to that of a simple
quantum mechanics problem in a non-trivial potential, and we can use
our knowledge about quantum mechanical scattering problems to
find good ways to solve the equation.

In linear theory, each Fourier mode will evolve independently. In
fact, we are intersted in following modes which early in the contracting
phase have a wavelength smaller than the Hubble radius
and then exit the Hubble radius at some point in time (i.e. the
Hubble radius decreases such that the wavelength becomes
larger). Thus, we Fourier transform the gauge field
\begin{eqnarray}\label{sol_A}
 \tilde{A}_\nu(\xi^\mu) \, = \, \int_{-\infty}^{\infty} d^3\vec{k} ~c_A(\vec{k}) \tilde{A}_{k}(t) \epsilon_\nu e^{i\vec{k}\cdot\vec{\xi}} ~,
\end{eqnarray}
where $\epsilon_\nu$ is the polarization unit vector. The Fourier mode 
$A_{\vec{k}}$ then obeys the following equation of motion: 
\begin{eqnarray} \label{kEoM}
 \ddot{\tilde{A}}_k + (k^2+M_{\rm YM}^2)\tilde{A}_k \, = \, 0  \, ,
\end{eqnarray}
which is a harmonic oscillator equation with time-dependent mass. Upon
quantization, the Fourier mode can be written as a combination of creation
and annihilation operators, and the time-dependence of the mass leads
to squeezing of the wave function in the same way that infrared modes
of cosmological perturbations and gravitational waves are squeezed
on super-Hubble scales in a dynamical cosmological background. 

The general solution of a Fourier mode of the gauge field can be
written in terms of Bessel functions:
\[
 \tilde{A}_k(t) \, = \,
\begin{cases}
 (-t)^{\frac{1}{2}} \bigg[ {\cal D}^-_J(k) J_{\nu_g}(-kt) 
 + {\cal D}^-_Y(k) Y_{\nu_g}(-kt) \bigg]~ & \quad \text{if} ~~~t \leq -\xi , \\
{\cal A} ~{\rm exp}[\beta t] + {\cal B} ~{\rm exp} [-\beta t] ~& \quad \text{if}~~~ -\xi \leq t \leq \xi, \\
{t}^{\frac{1}{2}} \bigg[ {\cal D}^+_J(k) J_{\nu_g}(kt) 
 + {\cal D}^+_Y(k) Y_{\nu_g}(kt) \bigg]~ & \quad \text{if}~~~ t \geq \xi \, ,
\end{cases}
\label{A_k_general}
\]
with the index 
\be
\nu_g \, = \, \frac{1+\alpha}{2}~~~~~~~\alpha = \sqrt{3} \, ,
\ee
and
\ben
\beta \equiv \sqrt{\frac{\alpha(\alpha+2)}{\xi^2}-k^2} \, .
\label{B-2}
\een
For $t < \xi$ the solution of the above mode equation involves two coefficients ${\cal D}^-_J$ and 
${\cal D}^-_Y$ which can be determined by matching the bulk solution and the
boundary operator at the surface of $- t_b$. 
We are interested in modes which start out in the vacuum state early during
the phase of contraction, i.e. in their Bunch-Davies \cite{BD} state. 

On sub-Hubble scales (large values of $kt$) both modes
are oscillating. For small values of $kt$ the modes have very
different asymptotics. If ${\cal D}^-_Y=0$ around the moment $- t_b$
then the asymptotic form of the solution is 
\begin{eqnarray}\label{sol_A_k_A}
  \tilde{A}_k(t) \, \sim \, {\cal D}^-_J \frac{{|t|}^{\frac{1}{2}}}{\Gamma_{1+\nu_g}}(\frac{|kt|}{2})^{\nu_g} ~,
\end{eqnarray}
i.e. it is in general a decaying mode as $t \rightarrow 0$. The second mode scales
as
\begin{eqnarray}\label{sol_A_k_B}
  \tilde{A}_k(t) \, \sim \, -{\cal D}^-_Y \frac{\Gamma_{\nu_g}}{\pi} {|t|}^{\frac{1}{2}}(\frac{2}{|kt|})^{\nu_g} ~,
\end{eqnarray}
which is a growing mode which in fact diverges as $t \rightarrow 0$.
This is a reflection of the fact that the mass term diverges.
The physical solution will be dominated by the growing mode
unless ${\cal D}_Y$ vanishes.  But it will not vanish in the general case. In
particular, if we were to match the solutions to a Bunch-Davies vacuum, then we
would expect the magnitude of both coefficients $D^-_{J,Y}$ in (\ref{A_k_general}) to be of the same order.

\subsection{Determination of the Spectrum of the Boundary Fluctuations at $-t_b$}

A key step in our analysis is to extract the spectrum of the
boundary gauge field from that of the bulk dilaton.
The coefficients ${\cal D}_J$ and ${\cal D}_Y$ are
determined via (\ref{Operator}) by taking the limit of the dilaton
fluctuations at the boundary. We make this identification at the time $- t_b$
when the coupling constant vanishes. At first sight, we are faced with a puzzle:
the right-hand side of (\ref{Operator}) is quadratic in the gauge
field, the left-hand side is linear in the dilaton. It is thus non-trivial
to infer the gauge field fluctuations from the bulk dilaton. The
approach we will take is to look for a power law form of the gauge
field Fourier modes $A_{\mu}(k)$ which yields the spectrum
of the bulk dilaton fluctuations we are starting with. 

The dominant mode for infrared modes in ${\rm Tr}[F^2(\xi)]$ is the term
${\dot A_i}^2$. Inserting the Fourier expansion of
$A_i(x)$ yields
\be
 {\dot A_i}^2 \, = \, \int d^3k_1 d^3k_2 
 {\dot A_i}(k_1) {\dot A_i}(k_2) e^{i(k_1 + k_2)x} V \, ,
 \ee
 where $V$ is the normalization volume used to define the Fourier
 transform. We introduce new momenta
 \ba
 k_1 \, &=& \, \frac{1}{2} (k + k^{\prime}) \, \nonumber \\
 k_2 \, &=& \, \frac{1}{2} (k - k^{\prime}) \, .
 \ea
 We write down the Fourier expansion of the dilaton field $\varphi$, and insert
 into (\ref{Operator}) and (\ref{A-1}) to identify Fourier coefficients. This leads to
 \be
 \varphi(k) \, = \, 
\frac{1}{4} \int d^3k^{\prime} {\dot A_i}(\frac{k + k^{\prime}}{2}) {\dot A_i}(\frac{k - k^{\prime}}{2})
V^{1/2}
\label{D-1}
\ee
where $\varphi$ is defined in (\ref{A-1}).

One way to find a consistent $A_i(k)$ to give rise to a power law bulk spectrum 
$\varphi (k) \sim k^{-\gamma}$ is as follows. We can divide this integral into a 
region $R_1$ with $k^{\prime} < k$ and a region
$R_2$ with $k^{\prime} > k$. In the integral over region $R_1$ we can set $k^{\prime} = 0$
to find the approximate result
\be
k^3 {\dot A}_{\mu}^2(k) \, \sim \, k^{-\gamma} 
\ee
and hence (making use of the fact that on super-Hubble scales $\dot{A} \sim H A$)
\be \label{scaling}
A_{i}(k) \, \sim \, k^{-\frac{\gamma+3}{2}} \, .
\ee
It is straightforward to check that this is a consistent solution for any $\gamma > 0$.  
Substituting (\ref{scaling}) into (\ref{D-1}) we get
\ben
\varphi (k) \sim \int d^3 k^\prime (k^2 - (k^\prime)^2)^{-(\gamma+3)} = k^{-\gamma} \int d^4 q (1-q^2)^{-\frac{\gamma + 3}{2}}
\een
The integral over $q$ is convergent when $\gamma > 0$. In particular, for a 
scale invariant spectrum $\gamma = 3/2$.


For cosmological scales we are interested in, the modes are outside
of the Hubble radius at the time $- t_b$. Hence, we can use the
small argument limit of the Bessel functions. We will assume that at $t =  - t_b$
both modes have the same amplitude. Let us denote the total amplitude
of $A(k)$ (we are dropping the index $i$) by $2 {\tilde{A}}$ (which is k-dependent).
Then we find the following values of the coefficients $D_{Y}^{-}$ and
$D_J^{-}$ before the bounce 
\ba 
D_{Y}^{-} \, = \, {\tilde{A}} |- t_b|^{-1/2} \bigl( \frac{| - k t_b|}{2} \bigr)^{\nu_g} \nonumber \\
D_{J}^{-} \, = \,  {\tilde{A}} |- t_b|^{-1/2} \bigl( \frac{| - k t_b|}{2} \bigr)^{-\nu_g} \, .
\label{Icoeffs}
\ea
 
At this point we know the equation of motion and the initial conditions
for the boundary gauge field at the time $- t_b$ when the boundary
theory becomes weakly coupled and when we begin the evolution of
the fluctuations on the boundary. The evolution can be followed by 
determining the coefficients $\cA, \cB, \cD^+_J,\cD^+_Y$ by standard 
matching of the function $\tA_k (t)$ and its time derivative at $t = \pm \xi$, 
as detailed in the next section. 

\subsection{Evolution of Boundary  Fluctuations through the Singularity}

It is useful to discuss matching of solutions of the general form
\[
 \tilde{A}_k(t) \, = \,
\begin{cases}
A_- (k) F_1(t) + B_-(k) G_1(t) ~ & \quad \text{if} ~~~t \leq -\xi , \\
{\cal A} ~{\rm exp}[\beta t] + {\cal B} ~{\rm exp} [-\beta t] ~& \quad \text{if}~~~ -\xi \leq t \leq \xi, \\
A_+ (k) F_2(t) + B_+(k) G_2(t)  & \quad \text{if}~~~ t \geq \xi .
\end{cases}
\label{mat-0}
\]
Matching $\tA$ and its time derivative across $t = \pm \xi$ then leads to 
\begin{eqnarray}
A_+ & = & \frac{1}{\Delta} \{ \cosh(2\beta\xi) [F_1(-\xi){\dot{G}}_2(\xi) - {\dot{F}}_1(-\xi)G_2(\xi)] +\sinh(2\beta\xi)[\frac{1}{\beta}{\dot{F}}_1(-\xi){\dot{G}}_2(\xi) -\beta F_1(-\xi) G_2(\xi) ] \} A_- \nonumber \\   
& & + \frac{1}{\Delta} \{ \cosh(2\beta\xi) [G_1(-\xi){\dot{G}}_2(\xi) - {\dot{G}}_1(-\xi)G_2(\xi)] +\sinh(2\beta\xi)[\frac{1}{\beta}{\dot{G}}_2(\xi){\dot{G}}_1(-\xi) -\beta G_1(-\xi) G_2(\xi) ] \} B_- 
\label{mat-1}
\end{eqnarray}
\begin{eqnarray}
B_+ & = & \frac{1}{\Delta} \{ \cosh(2\beta\xi) [F_2(\xi){\dot{F}}_1(-\xi) - {\dot{F}}_2(\xi)F_1(-\xi)] +\sinh(2\beta\xi)[-\frac{1}{\beta}{\dot{F}}_1(-\xi){\dot{F}}_2(\xi) +\beta F_1(-\xi) F_2(\xi) ] \} A_- \nonumber \\   
& & + \frac{1}{\Delta} \{ \cosh(2\beta\xi) [F_2(\xi){\dot{G}}_1(-\xi) - {\dot{F}}_2(\xi)G_1(-\xi)] +\sinh(2\beta\xi)[-\frac{1}{\beta}{\dot{F}}_2(\xi){\dot{G}}_1(-\xi) + \beta G_1(-\xi) F_2(\xi) ] \} B_- \, ,
\label{mat-2}
\end{eqnarray}
where we have defined
\ben
\Delta \equiv F_2(\xi) {\dot{G}}_2(\xi) - G_2(\xi) {\dot{F}}_2(\xi) \, .
\een
To specialize to the case of interest we need to substitute
\begin{eqnarray}
& F_1(t) \equiv (-t)^{1/2}J_{\nu_g}(-kt) &  ~~~~~~~F_2(t) \equiv (t)^{1/2}J_{\nu_g}(kt) \nonumber \\
& G_1(t) \equiv (-t)^{1/2}Y_{\nu_g}(-kt) &  ~~~~~~~G_2(t) \equiv (t)^{1/2}Y_{\nu_g}(kt) \nonumber \\
& A_\pm = \cD^\pm_J~~~~~&~~~~~~B_\pm = \cD^\pm_Y \, .
\label{mat-3}
\end{eqnarray}

Since we are always interested in wavenumbers small compared to the 
various time scales $t_b$ and $\xi$, we can replace the Bessel functions 
by their small argument values for $k\xi \ll 1$. This yields
\begin{eqnarray}
2\nu \cD^+_J & = & \bigg[-\cosh (2\beta \xi) + \sinh(2\beta\xi) \bigg( \frac{1}{4\beta\xi}(1-4\nu_g^2)+ \beta\xi \big) \bigg] \cD^-_J \nonumber \\
& + & \big(\frac{2}{k\xi} \big)^{2\nu_g} \bigg[ \cosh(2\beta\xi) (2\nu_g -1) + \sinh (2\beta\xi) \bigg(\beta\xi + \frac{1}{4\beta\xi}(1-2\nu_g)^2 \bigg) \bigg] \cD^-_Y \nonumber \\
2\nu \cD^+_Y & = & \big(\frac{k\xi}{2} \big)^{2\nu_g}\bigg[\cosh (2\beta \xi) (2\nu_g+1) - \sinh(2\beta\xi) \bigg( \frac{1}{4\beta\xi}(1+2\nu_g)^2+ \beta\xi \bigg) \bigg] \cD^-_J \nonumber \\
& + & \bigg[ \cosh(2\beta\xi) + \sinh (2\beta\xi) \bigg(\beta\xi + \frac{1}{4\beta\xi}(1-4\nu_g^2 \bigg) \bigg] \cD^-_Y
\, .
\label{mat-4}
\end{eqnarray}
We now substitute the initial conditions, (\ref{Icoeffs}) to calculate $\tA_k(t)$ at time $t=t_b$. 
The result is
\begin{eqnarray}
\tA_k (t_b) & = & \frac{\tA (k)}{2\nu_g} \bigg[2\sinh (2\beta\xi) \bigg( \beta \xi + \frac{1}{4\beta \xi}(1-4\nu_g^2) \bigg) \nonumber \\
& + & \bigg( \cosh (2\beta \xi) (2\nu_g+1) - \sinh(2\beta\xi) \big(\beta \xi + \frac{1}{4\beta \xi} ( 1+ 2\nu_g)^2 \big) \bigg) \bigg( \frac{\xi}{t_b} \bigg)^{2\nu_g} \nonumber \\
& + & \bigg( \cosh (2\beta \xi) (2\nu_g-1) + \sinh(2\beta\xi) \big(\beta \xi + \frac{1}{4\beta \xi} ( 1- 2\nu_g)^2 \big) \bigg) \bigg( \frac{t_b}{\xi} \bigg)^{2\nu_g} \bigg] \, .
\label{mat-5}
\end{eqnarray}
The significant point about (\ref{mat-5}) is that in the $k\xi \ll 1$ limit, the expression 
inside the square bracket becomes independent of the momentum $k$.
This is because $\beta = \bigg(\frac{\alpha(\alpha+2)}{4\xi^2}-k^2 \bigg)^{1/2}$, 
leading to $\beta\xi \approx \frac{1}{2}\sqrt{\alpha(\alpha+2)}$. The sole momentum 
dependence comes from the overall factor $\tA(k)$ which is the amplitude at $t = -t_b$.
This means that the spectrum of $\tA_k$ is the same as $t=-t_b$ and $t= t_b$. The 
amplitude is however amplified, since for $t_b \gg \xi$ we get
\ben
\tA_k(t_b) \approx \frac{1}{4\nu_g} \bigg[ \cosh (2\beta \xi) (2\nu_g-1) + \sinh(2\beta\xi) \bigg(\beta \xi + \frac{1}{4\beta \xi} ( 1- 2\nu_g)^2 \bigg) \bigg] \bigg( \frac{t_b}{\xi} \bigg)^{2\nu_g} \tA_k (-t_b) \sim \bigg( \frac{t_b}{\xi} \bigg)^{2\nu_g} \tA_k (-t_b) \, .
\label{mat-6}
\een
The amplification is due to squeezing of the perturbation modes while their wavelength 
is larger than the Hubble result.

This result holds for arbitrary functional form of $\tA(k)$. In particular, when $\tA(k)$ is a 
power law and given by a scale invariant spectrum at $t=-t_b$ the spectral index does 
not change at $t=t_b$. 

This is our main result. In the next section we will argue that this result in the boundary 
theory implies that the spectrum of cosmological fluctuations on a constant $z$ slice in 
the bulk is also unchanged as the system goes through $t=0$.

\subsection{The General Result}

From the above analysis it is now clear that the final result does not really depend 
on the details of the time dependence of the boundary coupling. This
conclusion agrees with what was found when studying non-singular
bulk cosmological models (see e.g. \cite{RHBmbRev} for a recent review). Consider the 
equation for the gauge field perturbation (\ref{B-0}) with a function $M_{YM}^2(t)$ 
which is smooth and bounded everywhere. Whenever the equation (\ref{kEoM}) has 
a regular solution for $k = 0$, it is clear that $\tA_i$ and its time dervative at $ t = t_b$ 
is related to those at $ t = -t_b$ by a Bogoliubov transformation 
$$\left[ \begin{array}{c} \tA_i(k,t_b)\\ {\dot{\tA}}_i(k,t_b)\end{array} \right] = \begin{bmatrix} M_{11} & M_{12} \\ M_{21} & M_{22} \end{bmatrix} \left[ \begin{array}{c}\tA_i(k,-t_b)\\ {\dot{\tA}}_i(k,-t_b)\end{array} \right]$$
where the matrix $M_{ij}$ depends on the potential, but not on $k$. Therefore 
if the initial conditions are such that 
\ben
\tA_i(k,-t_b) \sim {\dot{\tA}}_i(k,-t_b) \sim B(k)
\een
it follows that
\ben
\tA_i(k,t_b) \sim {\dot{\tA}}_i(k,t_b) \sim B(k)
\een
up to factors which do not depend on $k$. The specific power law enhancement 
factor in (\ref{mat-6}) is a feature of the fact that the coupling constant is a power law 
away from the region of small coupling.

Therefore the momentum dependence of the fields in the future is the same as 
that in the past for all momenta small compared to all other scales in the problem.

\section{Reconstruction of the Bulk from the Boundary Data}

Given the amplitude and spectrum of the gauge field fluctuations after
the singularity, we can reconstruct the late time spectrum of the
bulk dilaton after the bounce. We evolve the boundary data until
the time $+t_b$ when the 't Hooft coupling becomes unity.

As we emphasized above, the nature of the entire spacetime at positive 
times in the presence of these flutuations is not known and might contain 
a black brane in the bulk. We will, however, assume that there is a region 
near the boundary where the curvatures are small enough to enable us to 
use the AdS/CFT dictionary. In partcular we will assume that the bulk dilaton 
field at a point $(z,\vx,t)$ for small enough $z$ can be reconstructed from 
the dual boundary operator $\cO$ using a bulk boundary map of the form
\be \label{recon}
\varphi(z, \vx, t) \, = \, \int dt' d^3x' K(\vx', t' | z, \vx, t) {\cal O}(\vx', t') \, ,
\ee
where the kernel $K(x', t' | z, x, t)$ (where
$(t', x')$ are boundary coordinates and $(t, x, z)$ are
the bulk coordinates) is non-vanishing
only for points within the $AdS$ causal wedge (see Fig. 3) of
compact support. Due to the translation symmetry of the background,
we assume that the kernel $K$ depends only on the relative spatial
separation $ | \vec{x}? - \vec{x} | $.

The idea of the reconstruction is depicted in Figure 3: we
consider a wedge sticking into the bulk with base on
the boundary. The wedge is centered at the time $t_b$
We need the boundary points of the wedge to all have
time coordinate $t' > \xi$ - otherwise the wedge will
intersect the singular part of the bulk. This will limit
the distance from the boundary to which we can
reconstruct the bulk.

\begin{figure*}[t]
\begin{center}
\includegraphics[scale=0.4]{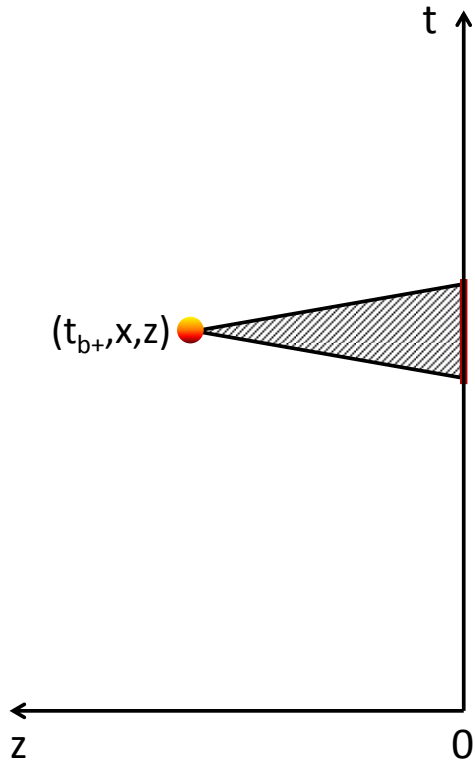}
\caption{Sketch of the reconstruction process for bulk operators
in the future of the singularity. The vertical axis is time,
the horizontal axis indicates the radial coordinate $z$.
The solid line is the boundary.The bulk field at a point with 
coordinates $(t_b^+, x, z)$ is given by integrating the
boundary data against the boundary-to-bulk propagator. The
integration involves data in the shaded region only. The solid curves
connecting the bulk point to the boundary are null geodesics.}
\end{center}
\end{figure*}

As we will see, we will in fact never need the
explicit form of the boundary-to-bulk propagator in order to extract
the spectrum of bulk dilaton fluctuations. The only property of this
propagator which will be used is that it is non-vanishing only
within the wedge shown in Figure 3. In particular as $z \rightarrow 0$, 
one recovers the correspondence (\ref{A-1}).

Now we turn to the extraction of the power spectrum of the bulk field
$\varphi$ in the future of the reconstruction time $t_b$. Our
starting point is equation (\ref{recon}) which gives the bulk field
in terms of the known boundary data. To extract the spectrum of
fluctuations, we need to take the Fourier transform of $\varphi(z, x, t)$
in the spatial hyperplane perpendicular to the $AdS$ radial coordinate
axis $z$:
\be
{\tilde \varphi}(z, k, t) \, = \, V^{-1/2} \int d^3x \varphi(z, x, t) e^{i k \cdot x}\, ,
\ee
where the tilde symbol indicates the Fourier transform. 

We now insert the reconstruction formula (\ref{recon}) into the
above. Introducing the new coordinates
\ba
{\bf x'} \, &=& \, {\bf x} + {\bf y'} \nonumber \\ 
t' \, &=& \, t + s \, .
\ea
the result becomes
\be
{\tilde \varphi}(z, k, t) \, = \, V^{-1/2} \int ds d^3y' K(y',s | z,t) e^{- i{\bf k} {\bf y'}}
\int d^3x e^{i {\bf k} ({\bf x} + {\bf y'})}  {\cal O}(x + y', t + s) \, .
\ee
The final integral simply gives the Fourier mode of the boundary
operator:
\be \label{almost}
{\tilde \varphi}(z, k, t) \, = \, \int ds d^3y' K(y', t + s | z, t) e^{- i{\bf k} {\bf y'}} {\cal O}(k, t+s) \, .
\ee
We are interested in values of $k$ which are small compared to both
the inverse Hubble radius and the $AdS$ radius. Assuming that
the propagator $K$ has support within the $AdS$ causal wedge
(see Figure 3) of the bulk point, then for all values of the boundary 
coordinate ${\bf y'}$ for which $K$ does not vanish we have 
${\bf k y'} \ll 1$, and one has approximately
\be \label{almost2}
{\tilde \varphi}(z, k, t) \, \simeq \, \int ds d^3y' K(y', s | z, t) {\cal O}(k, t+s) \, .
\ee

The $k$- dependence of the bulk fluctuation is therefore completely 
determined by the $k$- dependence of the expectation value of the 
operator $\cO$, which - in our framework - is in turn determined by the 
$k$- dependence of the gauge field fluctuations $A_k(t)$.  We have seen 
that the spectral index of the latter does not change when we cross 
the "singularity". The time integral in (\ref{almost2}) has an extent which 
is roughly of the same order as $z$. Therefore for $z$ small enough, the 
$k$ dependence of ${\tilde \varphi}(z, k, t)$ is basically given by the 
$k$-dependence of the gauge field fluctuations. Hence, we find that the
spectral index of fluctuations close enough to the boundary does not 
change when matching across the ``singular'' region (singular in quotation 
marks because we have cut off the actual singularity).
On the other hand, the amplitude
of the spectrum changes by a factor ${\cal F}$ given by
\be
{\cal F} \, = \, \bigl( \frac{t_b}{\xi} \bigr)^{2 \nu_g} \, .
\ee
In particular, the bulk spectrum on a small $z$ slice is chosen to be scale invariant, i.e.
$\varphi (k) \sim k^{-3/2}$ the spectrum continues to be scale invariant.

\section{Conclusion and Discussion}

In this paper we have studied the evolution of linearized test field
fluctuations in a deformed $AdS/CFT$ cosmology. The deformed
$AdS$ background which is the basis of our study is obtained
by introducing a nontrivial time-dependence for the dilaton and
choosing the metric such that the bulk supergravity equations of
motion are satisfied. The background begins with a contracting
phase which approaches a bulk singularity at time $t = 0$. As the singularity
is approached, the background becomes highly curved. At
the time $- t_b < 0$ the gravitational theory becomes strongly
coupled. However, the dual gauge theory which lives on the
boundary becomes weakly coupled. After the singularity, the
bulk expands, and after some time $t_b > 0$ the bulk theory
once again becomes weakly coupled.

Our goal is to compare the spectrum of fluctuations of the bulk
field in the far past ($t < - t_b$) and in the far future ($t > t_b$).
Specifically, we consider fluctuations in the dilaton field. We evolve
the fluctuations in the bulk until $t = - t_b$, map them onto
the boundary at that time, infer the spectrum of the boundary
gauge field fluctuations at this time and then evolve the
boundary gauge field fluctuations forward in time, past the
singularity, until the time $t = t_b$. At that time, we
reconstruct the bulk dilaton field using boundary-to-bulk
propagators which are nonvanishing only in a `$AdS$ causal wedge
wedge, and compute the spectrum of the fluctuations.

Since the boundary fluctuations blow up at the time $t = 0$ in
spite of the fact that the boundary theory is weakly coupled,
we need to smooth out the singularity. We, therefore, modify the 
dilaton between $-\xi < t < \xi$, where $\xi$ is a cutoff scale and 
then match the boundary fluctuations in a standard fashion.

Our main result is  that the spectral index of the dilaton fluctuations 
near the boundary is the same in the far past and the far future.  
While we do not have the tools to map out the future space-time, 
we believe that our result will have important implications for 
pre big-bang scenarios of cosmology.

\acknowledgements{We are grateful for discussion with many experts during the
long period when this project was in progress. In particular, we wish to thank
David Lowe, Omid Saremi, Misha Smolkin, Alberto Enciso, Niky Kamran, and Sandip Trivedi. 
We are grateful for support from the Banff International Research Station
where the work on this project was started. During the course of the
project we benefitted from support by grants from FQXi and from the
Mathematical Physics Laboratory of the CRM in Montreal. We gratefully
acknowledge support from the CRM during a recent workshop during
which this project was completed. The work
at McGill has been supported by an NSERC Discovery Grant, and by funds from the 
Canada Research Chair program. SD acknowledges support from the US National
Science Foundation under grants NSF-PHY-1214341 and NSF-PHY-1521045.
EF would like to thank CNPq (Science without Borders)  for financial support. 
YFC is supported in part by the National Youth 
Thousand Talents Program and by the USTC start-up funding under Grant No. KY2030000049. 
YW is supported by Grant HKUST4/CRF/13G issued by the Research Grants Council
(RGC) of Hong Kong. During the final writing of the manuscript RB was supported
by a Simons Foundation fellowship and by a Senior Fellowship at the Institute
for Theoretical Studies of the ETH Zurich, with support provided by 
Dr. Max R\"ossler, the Walter Haefner Foundation and the ETH Zurich Foundation.}

\vskip0.2cm

\appendix{\centerline{\bf Appendix A: Analysis in Terms of the Original Variables}}

\vskip0.2cm

The kinetic energy term of the gauge fields on the boundary is not
of canonical form. This was the motivation for rescaling the
gauge field. The rescaling factor, however, diverges at $t = 0$.
Hence, we might worry that the divergence of the rescaled gauge
field ${\tilde A}_{\mu}$ which we found is a result of this
rescaling, and that the evolution in terms of the original
variables $A_{\mu}$ might be better behaved.

In fact, the equation of motion for the fluctuation of
the original variable is
\be
\bigl( - \partial_t^2 + \frac{\sqrt{3}}{t} \partial_t + \partial_i \partial^i \bigr) A_j 
\, = \, 0 \, ,
\ee
which has Fourier mode solutions
\be
A_j(kt) \, = \, t^{\nu_g} \bigl( c_{+} J_{\nu_g}(kt) + c_{-} Y_{\nu_g}(kt) \bigr) \, .
\ee
The first mode goes to zero at $t = 0$ whereas the second mode
approaches a finite value. Hence, there is indeed no divergence
in the solutions. However, the first mode has a branch cut at
$t = 0$. Hence, matching conditions are still required in
order to evolve the solutions from negative to positive
values of $t$. 

In the spirit of the AdS/CFT correspondence it would be nice
not to have to impose any cutoffs in the matching calculation.
This could have been expected since the gauge theory becomes
free at $t = 0$. Indeed, there is a matching of the two modes
at $t = 0$ for which there is no particle production at all (in
terms of the rescaled variables this matching corresponds to
having the derivative of ${\tilde A_{mu}}$ match to its inverse
between $t = \xi$ and $t = - \xi$). This matching, however, does
not correspond to what is done in standard quantum mechanics
problems and it misses the particle production which is expected
on physical grounds. 

Hence, it appears that a matching prescription is needed.
An explicit calculation shows if the matching prescription
is taken to be the same as the one we used for
the rescaled variables, that then the matching calculation
calculation in terms of the original variables leads 
to the same result as that obtained using the rescaled field.

\end{document}